\documentclass[twocolumn]{aastex631}
\usepackage{newtxtext,newtxmath}
\usepackage{graphbox,graphicx}
\usepackage[flushleft]{threeparttable}
\usepackage[T1]{fontenc}
\usepackage{graphicx}	
\usepackage{amsmath}	

\begin{document}

\title{The Morphology and Kinematics of a Giant, Symmetric Nebula Around a Radio-Loud Quasar 3C\,57: Extended Rotating Gas or Biconical Outflows?}

\author[0000-0002-2662-9363]{Zhuoqi (Will) Liu}
\affiliation{Department of Astronomy, University of Michigan, 1085 S. University, Ann Arbor, MI 48109, USA}
\correspondingauthor{Zhuoqi (Will) Liu}
\email{zql@umich.edu}

\author[0000-0001-9487-8583]{Sean D. Johnson}
\affiliation{Department of Astronomy, University of Michigan, 1085 S. University, Ann Arbor, MI 48109, USA}

\author[0000-0002-0311-2812]{Jennifer~I-Hsiu Li}
\affiliation{Center for AstroPhysical Surveys, National Center for Supercomputing Applications, University of Illinois Urbana-Champaign, Urbana, IL, 61801, USA}
\affiliation{Michigan Institute for Data Science, University of Michigan, Ann Arbor, MI, 48109, USA}
\affiliation{Department of Astronomy, University of Michigan, Ann Arbor, MI, 48109, USA}

\author[0000-0002-2470-5756]{Benoît Epinat}
\affiliation{Aix-Marseille Univ, CNRS, CNES, LAM (Laboratoire d’Astrophysique de Marseille), Marseille, France}
\affiliation{Canada–France–Hawaii Telescope, 65-1238 Mamalahoa Highway, Kamuela, HI 96743, USA}

\author[0000-0002-8459-5413]{Gwen C. Rudie}
\affiliation{The Observatories of the Carnegie Institution for Science, 813 Santa Barbara Street, Pasadena, CA 91101, USA}

\author[0000-0002-6455-2491]{Ana Monreal-Ibero}
\affiliation{Leiden Observatory, Leiden University, PO Box 9513, 2300 RA Leiden, The Netherlands}

\author[0000-0001-5804-1428]{Sebastiano Cantalupo}
\affiliation{Department of Physics, University of Milan Bicocca, Piazza della Scienza 3, I-20126 Milano, Italy}

\author[0000-0002-2941-646X]{Zhijie Qu}
\affiliation{Department of Astronomy \& Astrophysics, The University of Chicago, 5640 S. Ellis Avenue, Chicago, IL 60637, USA}

\author[0000-0002-8739-3163]{Mandy C. Chen}
\affiliation{The Observatories of the Carnegie Institution for Science, 813 Santa Barbara Street, Pasadena, CA 91101, USA}
\affiliation{Cahill Center for Astronomy and Astrophysics, California Institute of Technology, Pasadena, CA 91125, USA}

\author[0000-0002-0417-1494]{Wolfram Kollatschny}
\affiliation{Institut für Astrophysik und Geophysik, Universität Göttingen, Friedrich-Hund Platz 1, D-37077 Göttingen, Germany}

\author[0000-0003-3938-8762]{Sowgat Muzahid}
\affiliation{Inter-University Centre for Astronomy and Astrophysics (IUCAA), Post Bag 4, Ganeshkhind, Pune 411 007, India}

\author[0000-0002-8739-3163]{Fakhri S. Zahedy}
\affiliation{Department of Physics, University of North Texas, Denton, TX 76201, USA}
\affiliation{The Observatories of the Carnegie Institution for Science, 813 Santa Barbara Street, Pasadena, CA 91101, USA}

\author[0000-0001-6846-9399]{Elise Kesler}
\affiliation{Department of Astronomy, University of Michigan, 1085 S. University, Ann Arbor, MI 48109, USA}

\author[0000-0002-9141-9792]{Nishant Mishra}
\affiliation{Department of Astronomy, University of Michigan, 1085 S. University, Ann Arbor, MI 48109, USA}



\begin{abstract}
Gas flows between galaxies and the CGM play a crucial role in galaxy evolution. When ionized by a quasar, these gas flows can be directly traced as giant nebulae. We present a study of a giant nebula around a radio-loud quasar, 3C\,57 at $z\approx0.672$. Observations from MUSE reveal that the nebula is elongated with a major axis of $70 \, \rm kpc$ and a minor axis of $40 \, \rm kpc$. The nebula displays an approximately symmetric blueshifted-redshifted pattern along the major axis and multi-component emission features in its $\rm[O\,II]$ and $\rm [O\,III]$ profiles. The morphology and kinematics can be explained as rotating gas or biconical outflow, both of which qualitatively reproduce the observed position-velocity diagram. The 3C\,57 nebula is significantly more kinematically disturbed, with $\rm W_{80}$ (the line width encompassing 80\% of the flux) of approximately $300{-}400\,\rm km\,s^{-1}$, compared to $\rm H\,I$ gas in local early-type galaxies, which typically shows $\rm W_{80} \approx 50\,\rm km\,s^{-1}$. This velocity dispersion is comparable to the gas in cool-core clusters despite originating in a group 100 times less massive. For biconical outflow models, the inferred $10{-}20^{\circ}$ inclination angle is in tension with the unobscured nature of the quasar, as the dusty torus is expected to be perpendicular to the outflow. Neither a quiescent rotating gas origin nor an biconical outflow fully reproduces the observed kinematics and morphology of the 3C\,57 nebula, suggesting a more intricate origin likely involving both rotation and AGN feedback.
\end{abstract}
\keywords{quasars: supermassive black holes -- galaxies: groups -- intergalactic medium}


\section{Introduction} \label{sec:intro}
The baryon cycle, encompassing both gas inflows and outflows between galaxies and the circumgalactic/intergalactic medium (CGM/IGM), is central to understanding galaxy evolution. Inflows supply galaxies with fresh material, driving star formation and black hole growth \citep[e.g.,][]{1997ApJ...477..765C, 2013ApJ...768...74T}, while outflows, driven by processes such as Active Galactic Nuclei (AGN) and stellar outflows \citep[for a review, see][]{2012ARA&A..50..455F, 2018Galax...6..114Z}, tidal interactions \citep[e.g.,][]{2016MNRAS.461.2630M}, and ram pressure stripping \citep[e.g.,][]{2006ApJ...647..910H}, return enriched gas to the CGM/IGM. These feedback mechanisms also expel heavy elements produced by stars and supernovae from the ISM into the CGM/IGM to reproduce the observed mass–metallicity relation \citep[e.g.,][]{2004ApJ...613..898T, 2016MNRAS.456.2140M}. Consequently, studying the gas exchange around galaxies offers valuable insights into the processes shaping galaxy formation and evolution, and it is ranked as a key, long-term priority by the the 2020 Decadal Survey \citep{2021pdaa.book.....N}. 

Large scale gas reservoirs can be directly observed through H\,I 21-cm emission in the local Universe. However, observations of 21-cm emission from diffuse gas reservoirs are not currently possible beyond the local Universe. Most CGM observations in the distant Universe therefore rely on sensitive absorption spectroscopy of UV-bright background sources passing through the halos of foreground galaxies \citep[for a review, see][]{2017ARA&A..55..389T}. This approach, however, has inherent limitations in constraining the morphology and spatially resolved kinematics of gas flows except in rare cases \citep{2014MNRAS.438.1435C, 2018Natur.554..493L}. Alternatively, gas flows around galaxies, along with their morphology and kinematics, can be directly traced by observations of giant nebulae using cutting-edge wide-field integral field spectrographs (IFS) such as the Multi-Unit Spectroscopic Explorer (MUSE; \citealt{2010SPIE.7735E..08B}). Deep, coadded MUSE observations enable the detection of CGM emission around galaxies in Ly$\alpha$, $\rm [O\,II]$, $\rm Mg\,II$, and $\rm Si\,II$ \citep[e.g.,][]{2016A&A...587A..98W, 2023MNRAS.522..535D, 2023Natur.624...53G, 2024A&A...688A..37G, 2024arXiv240604399K}. However, such observations typically require tens to hundreds of hours of integration time to achieve the necessary sensitivity, making them resource-intensive.

Systems such as quasars substantially elevate the local ionizing radiation background, which increases the ionization rate and consequently the recombination rate under photoionization equilibrium. \textcolor{black}{This also boosts the hydrogen ionization fraction and the number of metal ions such as $\rm O^{+}$ and $\rm O^{2+}$ in the surrounding IGM/CGM, leading an amplified emission in recombination and collisionally excited lines.} At $z > 2$, systematic IFS surveys revealed ubiquitous giant H\,I Ly$\alpha$ nebulae around radio-quiet quasars extending over $100\, \rm kpc$ \citep[e.g.,][]{2014Natur.506...63C, 2016ApJ...831...39B, 2019ApJS..245...23C, 2020ApJ...894....3O, 2021MNRAS.503.3044F, 2021MNRAS.502..494M}. At $z<1.5$, the capabilities of IFS have led to discoveries of giant nebulae around quasars emitting in $\rm [O\,II]$, $\rm H\beta$, and $\rm [O\,III]$ \citep[e.g.,][]{2018ApJ...869L...1J, 2022ApJ...940L..40J, 2024ApJ...966..218J, 2021MNRAS.505.5497H, 2024MNRAS.527.5429L}. 

The Cosmic Ultraviolet Baryon Surveys (CUBS; \citealt{2020MNRAS.497..498C}) and the MUSE Quasar Blind Emitters Survey (MUSEQuBES; e.g., \citealt{2024MNRAS.528.3745D}) were designed to study CGM and IGM surrounding galaxies at $z\approx 0.1-1.4$, leveraging high-quality COS absorption spectra of UV luminous quasars. Such investigations necessitate comprehensive galaxy redshift surveys, achievable with deep MUSE observations. As a result, CUBS and MUSEQuBES acquired 30 deep MUSE datacubes around UV luminous, unobscured quasars at $z = 0.4-1.4$. These data serendipitously enable observations of rest-optical emission around the quasars. \citet{2024ApJ...966..218J} recently reported the frequent detection of large ionized circumgalactic nebulae around UV luminous quasars in these fields. Some of the nebulae exhibit irregular or filamentary morphologies, consistent with interactions and accretion, while others exhibit more regular morphologies. Comprehensive studies exploring the nebulae with irregular and filamentary morphologies have been published \citep[e.g.,][]{2018ApJ...869L...1J, 2022ApJ...940L..40J, 2021MNRAS.505.5497H, 2024MNRAS.527.5429L}, but no case studies have yet focused on nebulae with regular morphologies and kinematics. In this paper, we present the first case study of a nebula with a regular morphology and kinematics resembling those of rotation or biconical outflows around a radio-loud quasar at $z <1$, 3C\,57 in the CUBS+MUSEQuBES survey.

Throughout the paper, we adopt a flat $\Lambda$ cosmology with $\Omega_{\rm m}=0.3$, $\Omega_{\rm \Lambda}=0.7$, and $H_{0} = 70 \, \rm km \, s^{-1} Mpc^{-1}$. All magnitudes are given in the AB system \citep{1983ApJ...266..713O}, unless otherwise stated. We use proper kpc (pkpc) as the unit of distance.

\section{Observations and Data} \label{sec:OD}
The quasar 3C\,57 hosts a $\approx 70$ kpc scale circumgalactic nebula, discovered as one of larger nebulae around 30 UV luminous quasars at $z<1$ \citep{2024ApJ...966..218J}. To investigate this nebula, we observed the quasar field around 3C\,57 using MUSE on the Very Large Telescope (VLT) as part of MUSEQuBES survey (PI: J. Schaye, PID: 094.A-0131(B) \& 096.A-0222(A)). We obtained four exposures collected between August 23rd, 2015 and October 11th, 2015 with a total exposure time of 2.0 hr with median seeing full-width-at-half-maximum (FWHM) conditions of $0.7''$. We then reduced the MUSE data using three independent pipelines: CubEx \citep{2019MNRAS.483.5188C}, the MUSE GTO team pipeline \citep{2014ASPC..485..451W}, and the ESO reduction pipeline \citep{2012SPIE.8451E..0BW}. All three pipelines produced consistent results, though with some differences in illumination corrections and night-sky-subtraction \citep[for more details, see][]{2020MNRAS.491.2057L, 2024ApJ...966..218J}. For simplicity, we converted the air wavelengths provided by the three pipelines to vacuum wavelengths. 

To study faint extended emission, we performed quasar light subtraction to remove both continuum and line emission from the spatially unresolved narrow-line and broad-line regions in the nucleus, following the method described in \citet{2018ApJ...869L...1J} and \citet{2021MNRAS.505.5497H}. This technique leverages the differences in spectral energy distribution between quasars and galaxies instead of relying on PSF measurements. Previous applications have demonstrated its effectiveness in eliminating spatially unresolved continuum, broad-line emissions, and narrow-line emissions from the nucleus.

We also obtained an image from the Advanced Camera for Surveys (ACS) on the \textit{Hubble Space Telescope} (\textit{HST}) to enable more sensitive and higher angular resolution characterization of galaxies in the quasar field. The image was taken in the F814W filter (PI: L. Straka, PID: 14660) with an exposure time of $2179$ seconds. We obtained the reduced image from the Barbara A. Mikulski Archive for Space Telescopes (MAST). To ensure uniformity in astrometry, we aligned the MUSE datacube and HST image with Gaia astrometry system using \texttt{Astrometry v1.5} \citep{2022zndo...6462441W}.

\section{Results} \label{sec:ME}
\subsection{Quasar properties}
3C\,57 is a luminous, radio-loud quasar \citep{2006A&A...455..773V}. To estimate its redshift, luminosity, and black hole mass, we extracted the MUSE spectrum with \texttt{MPDAF} \citep{2016ascl.soft11003B} before performing quasar and continuum subtraction. We chose an extraction aperture of $r=3''$, and measured the systemic redshift of the quasar by fitting the $\rm [O \, II] \lambda \lambda 3727, 3729$ doublet with a Gaussian profile. Following \citet{2010MNRAS.405.2302H}, we assumed a centroid of $\lambda3728.6$ for the doublet, expected from a $0.8{-}0.9{:}1$ doublet ratio. \textcolor{black}{We found $z = 0.6718 \pm 0.0002$, where the uncertainty reflects the scatter between the $\rm [O \, II]$ centroid and stellar absorption lines of SDSS quasars at similar redshift \citep{2010MNRAS.405.2302H}.}

In addition, we fit the extracted quasar's spectrum with the Python QSO fitting code (\texttt{PyQSOFit}; \citealt{2019MNRAS.482.3288G}) to estimate the bolometric luminosity and the black hole mass of 3C\,57. \texttt{PyQSOFit} models a quasar's spectrum with a combination of a power-law continuum, $\mathrm{Fe \, II}$ template, and sets of Gaussian line profiles for both the broad- and narrow-lines. From the fit, we computed a monochromatic luminosity at $5100$\AA \, of $\lambda L_{5100} \approx 6.9 \times 10^{45} \rm\ erg \, s^{-1}$ and a bolometric luminosity of $L_{\rm bol} \approx 6.3 \times 10^{46} \, \rm erg \, s^{-1}$ using the bolometric correction factor from \cite{2006ApJS..166..470R}. Finally, we determined a FWHM of $\rm H\beta$ as $3900 \, \rm km \, s^{-1}$ and used the corresponding single-epoch virial theorem relations from \cite{2006ApJ...641..689V} to infer a black hole mass of $M_{\rm BH}\approx 10^{8.9}\ {\rm M_{\odot}}$. According to \citet{2013ARA&A..51..511K}, this black hole mass corresponds to a stellar mass of $M_{*}\approx 10^{11.2}\ {\rm M_{\odot}}$ for the host galaxy, though this stellar mass may have a $>\!0.5$ dex uncertainty due to the systematic uncertainty in single-epoch black hole mass estimates and the scatter in the black hole mass-stellar mass relations. 

\subsection{The Group Environment of 3C\,57} \label{GE}
Radio-loud quasars typically reside in overdense environments with a halo mass of $\approx 10^{13} \rm M_{\odot}$, as determined from clustering studies \citep{2009ApJ...697.1656S}. Interactions between galaxies can strip gas from a galaxy's ISM/CGM, and lead to the formation of giant quasar nebulae when this gas falls within the quasar's ionization cone \citep{2018ApJ...869L...1J, 2024MNRAS.527.5429L}. Therefore, characterizing the group environment of the quasar is crucial before investigating the properties of the nebula.

To study the galactic environment of 3C\,57, we conducted a galaxy survey on 3C\,57, as detailed in \citet{2021MNRAS.505.5497H} and \citet{ 2024MNRAS.527.5429L}. In brief, we started by identifying continuum objects in both MUSE white light and the ACS$+$F814W images. We then fit each identified object with a linear combination of SDSS galaxy eigenspectra \citep{2012AJ....144..144B} to obtain the galaxy redshift. We computed the best-fit linear combination on a grid ranging from $z=0$ to $z=1$ with a step size of $\Delta z = 0.0001$, recording the goodness-of-fit statistic ($\chi^2$) over the entire grid. We adopted the redshift with the minimum global $\chi^2$ as our initial solution and then visually inspected each best-fit model to ensure robustness. For galaxies with both emission and absorption lines, we masked out strong emission lines and measured the redshift based on stellar absorption features when possible. This approach helps avoid potential bias in redshift measurements from large-scale nebulae in the field that may not be closely associated with the galaxies in question. From previous applications, the typical uncertainty in redshift measurements is $\sigma \approx \rm 20 \, km \, s^{-1}$ \citep{2021MNRAS.505.5497H}.

In the MUSE field of 3C\,57, we identified six galaxies including the quasar host, with LOS velocities $|\Delta v| < \rm 1500 \, km \, s^{-1}$ of the quasar systemic velocity. Outside MUSE FoV, ongoing wide-field spectroscopy follow-up with Magellan does not identify any potential group members within $ 500 \, \rm kpc$ from the quasar. Four of the six associated galaxies are more than $75 \, \rm kpc$ away from the quasar, and three of them are actively star-forming. One galaxy, G2, is near the quasar centroid, and its redshift can only be measured from emission lines. Additionally, we detected a galaxy G1 (not included as one of the six galaxies), which might be at the quasar's systemic redshift. However, its redshift cannot be determined due to its proximity to the quasar. Compared to other quasars with similar UV luminosity and MUSE survey depth, the number of group members around 3C\,57 is slightly below average, but remains within the expected range for radio-loud systems, which typically host $3{-}23$ galaxies \citep{2024ApJ...965..143L}, though this estimate is derived from a limited sample.

For the six group galaxies, we computed a mean LOS velocity of $\Delta v_{\rm group}=100\, \rm km\,s^{-1}$ relative to the quasar's systemic velocity. Due to the fact that the sample standard deviation estimator is statistically biased in the small $N_{\rm gal}$ regime, we calculated a bias-corrected velocity dispersion of $\sigma_{\rm group}=200\, \rm km\,s^{-1}$ following \citet{2020A&A...641A..41F}. We then estimated a halo mass of $\log(M_{\rm halo}/\rm M_{\odot}) \approx 12.7$ using Equation (1) from \cite{2013MNRAS.430.2638M}. In contrast, UV luminous quasars with more than 5 group members typically inhabit halos with masses of $\log(M_{\rm halo}/\rm M_{\odot}) \approx 13.5\pm 0.6$, and radio-loud quasars are generally found in even larger halos, with masses of $\log(M_{\rm halo}/\rm M_{\odot}) \approx 13.5{-}14.6$ \citep{2024ApJ...965..143L}. This suggests that 3C\,57 resides in a less massive halo compared to typical UV luminous quasar systems.

\subsection{Giant Nebula Characteristics} \label{NE}
Ionizing radiation from the accretion disks of AGN can ionize the gas to large distances, causing it to emit in recombination and collisionally-excited lines such as $\rm [O\,II]$, $\rm H\beta$, and $\rm [O\,III]$. As a result, wide-field IFS observations of quasar fields often reveal large nebulae. In this paper, we adopted the same methodology as \citet{2024ApJ...966..218J}, who presented the discovery of 3C\,57 nebula after conducting quasar light subtraction, continuum subtraction, and optimal extraction. We performed continuum subtraction locally for the $\rm [O\,II]$ and $\rm [O\,III]$ emission lines around the quasar, as described in \citet{2024MNRAS.527.5429L}. We first fit the continuum to the blue and red sides of these lines after masking the spectral region within $\pm500{-}1000\, \rm km \, s^{-1}$ of the expected observed wavelength at the quasar’s redshift. We then conducted optimal extraction with 3D segmentation, as described in \citet{2024ApJ...966..218J}. 

\begin{figure*}
    \centering
    \includegraphics[scale=0.8]{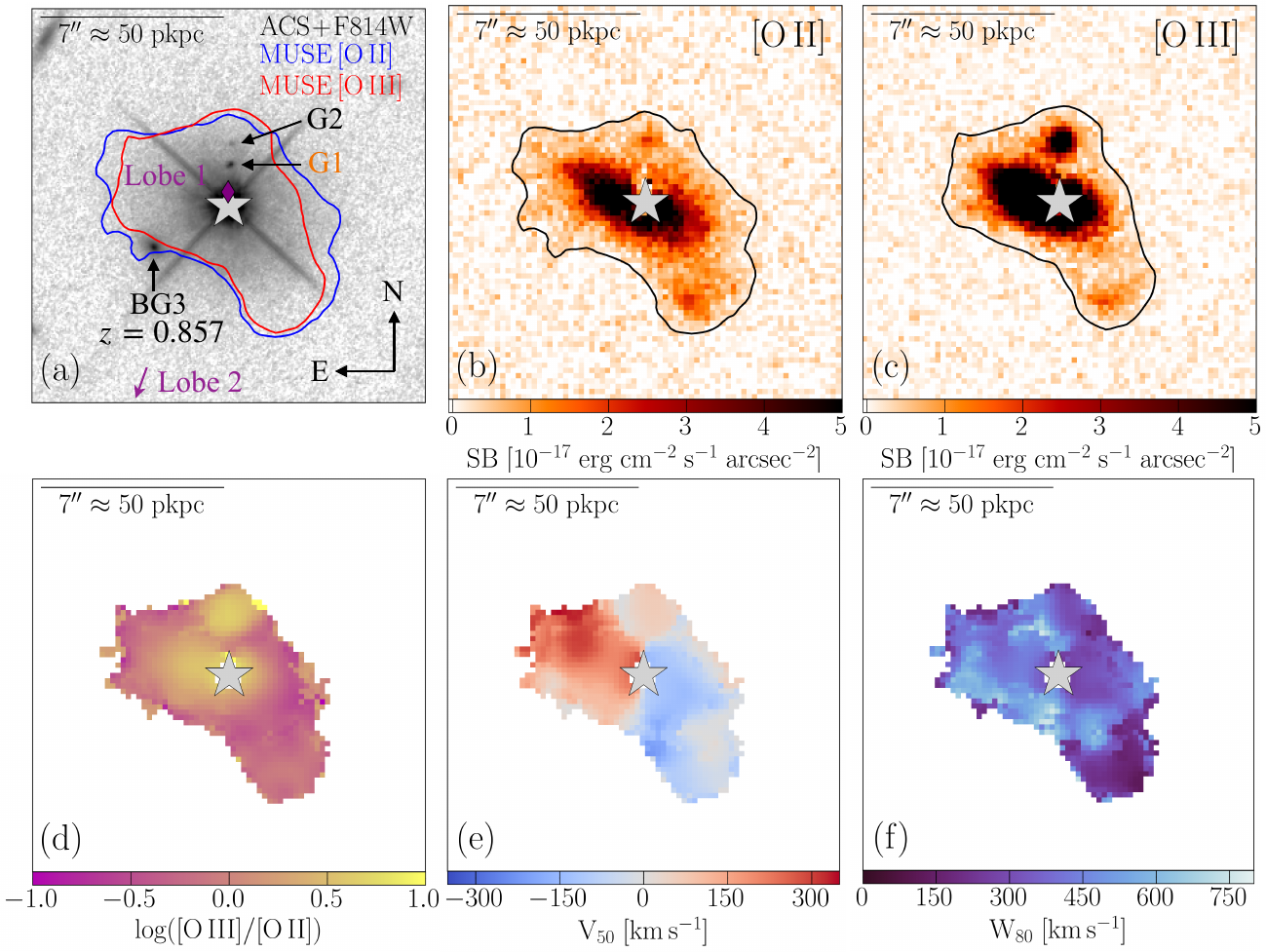}
    \caption{Visualizations of the nebula around 3C\,57. Panel (a): HST ACS+F814W image of the field, with galaxies labelled with their IDs. \textcolor{black}{The centroid of one radio lobe is marked with a purple diamond, while the second radio lobe lies outside the FoV and is indicated with a purple arrow.} Panel(b) and Panel (c): narrow-band $\rm [O \, II]$ and $\rm [O \, III]$ surface brightness maps generated from optimal extraction. These maps are overlaid with $\rm [O \, II]$ and $\rm [O \, III]$ surface brightness contours at levels of $0.2 \times 10^{-17} \rm \, erg \, cm^{-2} \, s^{-1} \, arcsec^{-2}$. The contours shown in panel (b) and panel (c) are overlaid on the HST image in blue and red respectively. Panel (d): map of nebular photoionization shown as the line ratio $\rm [O\,III]\lambda5008 / [O\,II]\lambda\lambda3727+3729$. Panel (e) and Panel (f): maps of the nebular $\rm V_{50}$ and $\rm W_{80}$.}
    \label{fig:SBs}
\end{figure*}

\textcolor{black}{This 3D segmentation technique is widely applied in the detection of giant Ly$\alpha$ nebulae at higher redshift \citep{2016ApJ...831...39B, 2019A&A...631A..18A, 2021ApJ...923..252S}. In summary, we first smoothed the datacube in both the spatial and spectral dimensions with a $\sigma=1.5$ pixels Gaussian kernel, and created a $S/N$ datacube. For each wavelength slice, we identified spaxels with a threshold of $S/N > 1.5$ in the smoothed $S/N$ datacube, connected adjacent spaxels above this threshold, and required at least 10 connected spaxels to define a detection. We began with the detection having the largest connected area, which defined its spatial segmentation. We then identified the wavelength slice where this largest connected area is. For spectral segmentation, we started at this layer and identified the continuous wavelength range (in both increasing and decreasing directions) meeting the threshold for each spaxel in the spatial segmentation, stopping when the condition was no longer met. We repeated this process for the detection with the second largest area after masking the previous detection, continuing until no additional detections remained.} We then integrated the unsmoothed flux in each spaxel over the spectral interval defined by the 3D segmentation to create surface brightness maps. For pixels falling below the S/N threshold, we used a background level of three spectral pixels at the wavelength where the mean S/N per pixel of the nebula is highest. The resulting $\rm [O\,II]$ and $\rm [O\,III]$ surface brightness maps for 3C\,57 are shown in Figure \ref{fig:SBs}, revealing a giant $\approx 70 \rm \, kpc$ nebula. 

To quantify the kinematics of the nebula, we jointly fit Gaussian line profiles to the quasar and continuum subtracted $\rm [O \, II]$ and $\rm [O \, III]$ datacubes. To enhance S/N, we first smoothed the data with a $\sigma=0.3''$ Gaussian kernel, which we choose to match the seeing disk. We incorporated the segmentation map from the optimal extraction to define the spatial fitting region. We initially fit a single kinematic component to the $\rm [O\,II]$ and $\rm [O\,III]$ emission lines in every spaxel of the nebula. Specifically, we jointly fit two Gaussian profiles for the $\rm [O\,II]$ doublet and one Gaussian profile for $\rm [O\,III]$, with shared redshift and velocity dispersion parameters across all lines. In most cases, the kinematic parameters shared by $\rm [O\,II]$ and $\rm [O\,III]$ result in good fits to the data, but in spaxels where the line profiles of $\rm [O\,II]$ and $\rm [O\,III]$ are inconsistent, we add an additional component to $\rm [O\,III]$ with significant flux, while this component contributes negligible flux in $\rm [O\,II]$. \textcolor{black}{We then adjusted the number of components based on the fit quality, evaluating both $\chi^2$ and the Bayesian Information Criterion (BIC). If adding more components led to a lower BIC, we adopted the solution with the higher number of components \citep[for more information on this approach, see][]{2007MNRAS.377L..74L}.} For majority of the spaxels in the nebula, one or two Gaussians are adequate to describe the line profile, though three components are needed in some regions. We note that complex kinematic features are easier to detect for spaxels with higher S/N. Therefore, we are more likely to miss line asymmetry at the edge of the nebula due to lower S/N. Finally, we visually inspected all spaxels to check if the best-fit model sufficiently reproduces the data. To visualize the kinematics of the nebula and compare with the literature \citep[e.g.,][]{2013MNRAS.436.2576L}, we produced $\rm V_{50}$ and $\rm W_{80}$ maps, which visualize the line center and width. In brief, we measured velocities at which a fraction of the model line flux is accumulated. $\rm V_{50}$ corresponds to the velocity at 50\% of the cumulative flux of the fitted Gaussian models. $\rm W_{80}$ represents the line width between the velocity range at 10\% and 90\% of the cumulative flux. We used the $\rm [O\,III]$ line profiles to construct these maps whenever it is possible, and used $\rm[O\,II]$ when $\rm [O\,III]$ is not detected. \textcolor{black}{For most spaxels with detections of both lines, $\rm [O\,II]$ and $\rm [O\,III]$ yield consistent $\rm V_{50}$ and $\rm W_{80}$ measurements.} The resulting $\rm V_{50}$, and $\rm W_{80}$ maps are displayed in panel (e) and (f) of Figure \ref{fig:SBs} respectively. 

To study the ionization state of the nebula, we also produced a $\rm [O \, III]/[O \, II]$ line ratio map by taking the ratio of the integrated fluxes of these lines. For spaxels not detected in $\rm [O \, III]$, we computed $3\sigma$ upper limits using $\rm [O \, III]$ error arrays by summing over the spectral dimension within a range determined by the $\rm [O \, II]$ optimal extraction. This map provides an opportunity to study the spatial dependence and distribution of the kinematics and the ionization state of the gas. \textcolor{black}{The $\rm [O \, III]/[O \, II]$ (O32) line ratio map is shown in panel (d). To further characterize the ionization state, we computed the geometric median, defined as the median value across all spaxels, and found $\log[\rm O32]\approx -0.05$. Since this value is dominated by less luminous regions and is not representative of the entire nebula, we also estimated a cumulative O32 by dividing the total $\rm [O \, III]$ flux by the total $\rm [O \, II]$ flux across the entire nebula, yielding $\log[\rm O32]\approx 0.20$. These values are higher than that of typical massive star-forming galaxies (median $\log[\rm O32]\approx-0.40$; \citealt{2018A&A...618A..40P}).}

\textcolor{black}{To investigate the ionization source of the nebula, we estimated an approximate $\log([\rm O\,III]/H\beta) \approx 0.7$. This suggests the nebula is ionization-bounded (large optical depth), as a matter-bounded cloud would exhibit a significantly lower $\log([\rm O\,III]/H\beta)$ due to oxygen being ionized beyond the $\rm O^{2+}$ state, while $\rm H\beta$, as a recombination line, remains unaffected \citep{2013MNRAS.430.2327L, 2024MNRAS.527.5429L}. Additionally, we found higher ionization state lines, including $\rm [Ne\,III]\lambda 3869$ and $\rm [Ne\, V]\lambda3427$ in the inner region of the nebula. The ionization potential for $\rm Ne^{4+}$ is 97 eV \citep{2011piim.book.....D}, far exceeding the 54 eV required to produce $\rm He^{2+}$, the highest energy photons typically generated in large numbers by stellar populations. Moreover, quasars with similar UV luminosity have been observed to ionize nebulae over large distances, as seen in \citet{2024MNRAS.527.5429L}. These factors support photoionization by the quasar \citep[e.g.,][]{2004ApJS..153....9G}, essentially ruling out star formation as the primary ionization source for most of the nebula. However, the observed line ratios could also be explained by fast shocks, as the nebular kinematics align with the velocities required to produce them \citep{2024MNRAS.527.5429L}. Thus, the ionization mechanism remains ambiguous, with both quasar photoionization and fast shocks as viable explanations.} To determine if dust affects these line ratios, we estimate the Balmer-line ratios at $\approx 4 \, \rm kpc$ ($0.5$ arcsec) east of the quasar centroid, where the surface brightness is the highest. With an extraction radius of $\approx 7 \, \rm kpc$ ($1$ arcsec), we find $\rm H\gamma/H\beta \approx 0.48 \pm 0.08$, consistent with Case B recombination \citep{2006agna.book.....O} in the absence of dust.

To examine the nebula and any relationship with galaxies in the quasar environment, we overlaid $\rm [O \, II]$ and $\rm [O \, III]$ emission contours over the HST image in the panel (a) of Figure \ref{fig:SBs}. We detected three galaxies enclosed or partly enclosed by the nebula. Galaxy G1 is spatially aligned with the northern part of the nebula, but its redshift remains undetermined due to insufficient S/N after quasar light subtraction. G2 corresponds to the bright knot in $\rm [O \, III]$, indicating its spatial coincidence with the nebula. BG3 on the southeast side is a background galaxy at a redshift of $z\approx0.857$ \textcolor{black}{determined from stellar absorption lines detected in the MUSE spectrum} and therefore unassociated with the nebula. Overall, the majority of the nebula does not closely align with nearby galaxies, suggesting that it is not arising from on-going galaxy interactions. 

The radio-loud quasar 3C\,57 exhibits two prominent radio lobes \citep{1993MNRAS.263.1023M, 1999ApJS..124..285R}. One lobe is centered near the quasar, while the other is located $15''$ southeast, corresponding to $100$ kpc at the quasar's redshift. To illustrate their positions relative to the quasar, we overlay the VLASS radio lobe positions on the HST image in the panel (a) of Figure \ref{fig:SBs} \citep{2020PASP..132c5001L}. The central lobe is marked with a purple diamond, while the southeastern lobe, located outside the FoV, is indicated by a purple arrow. The 3C\,57 nebula does not extend to the southeast radio lobe, but spatially overlaps with and surrounds the other radio lobe near the quasar's centroid, at last in projection.

The giant nebula surrounds the quasar with projected radii of $d \approx 10$ to $30\rm \, kpc$, and exhibits LOS velocities of $ \rm V_{50} \approx -250 \ \rm to \ {+}250 \ km \, s^{-1}$. Notably, the nebula displays a fairly symmetric blueshifted-redshifted pattern from the north-east to the south-west along the major axis. This kind of symmetric blueshifted-redshifted pattern can be produced by approximately edge-on rotation or biconical outflows. However, these two scenarios would produce distinct velocity dispersion maps. Gas in rotating disks typically shows more quiescent kinematics with $\sigma \approx 25{-}100 \, \rm km\,s^{-1}$,  or equivalently $\rm W_{80}\approx 60{-}250\, km\,s^{-1}$ (See discussion), while gas disturbed by outflows typical have $\rm W_{80}> 400\, km\,s^{-1}$ \citep{2016A&A...594A..44H}. The $\rm W_{80}$ map of 3C\,57 aligns more with the characteristics expected from outflows, showing velocity dispersions ranging from $\rm W_{80} \approx 300 \ \rm to \ 700 \ km \, s^{-1}$, with peaks in several spots over the northern and southern extensions. \textcolor{black}{An exception is the southwestern edge of the nebula, which exhibits quiescent kinematics. This region is kinematically distinct, with a $\rm W_{80}$ of $150 \, \rm km\,s^{-1}$, significantly lower than the rest of the nebula.}

\begin{figure*}
    \centering
    \includegraphics[scale=0.7]{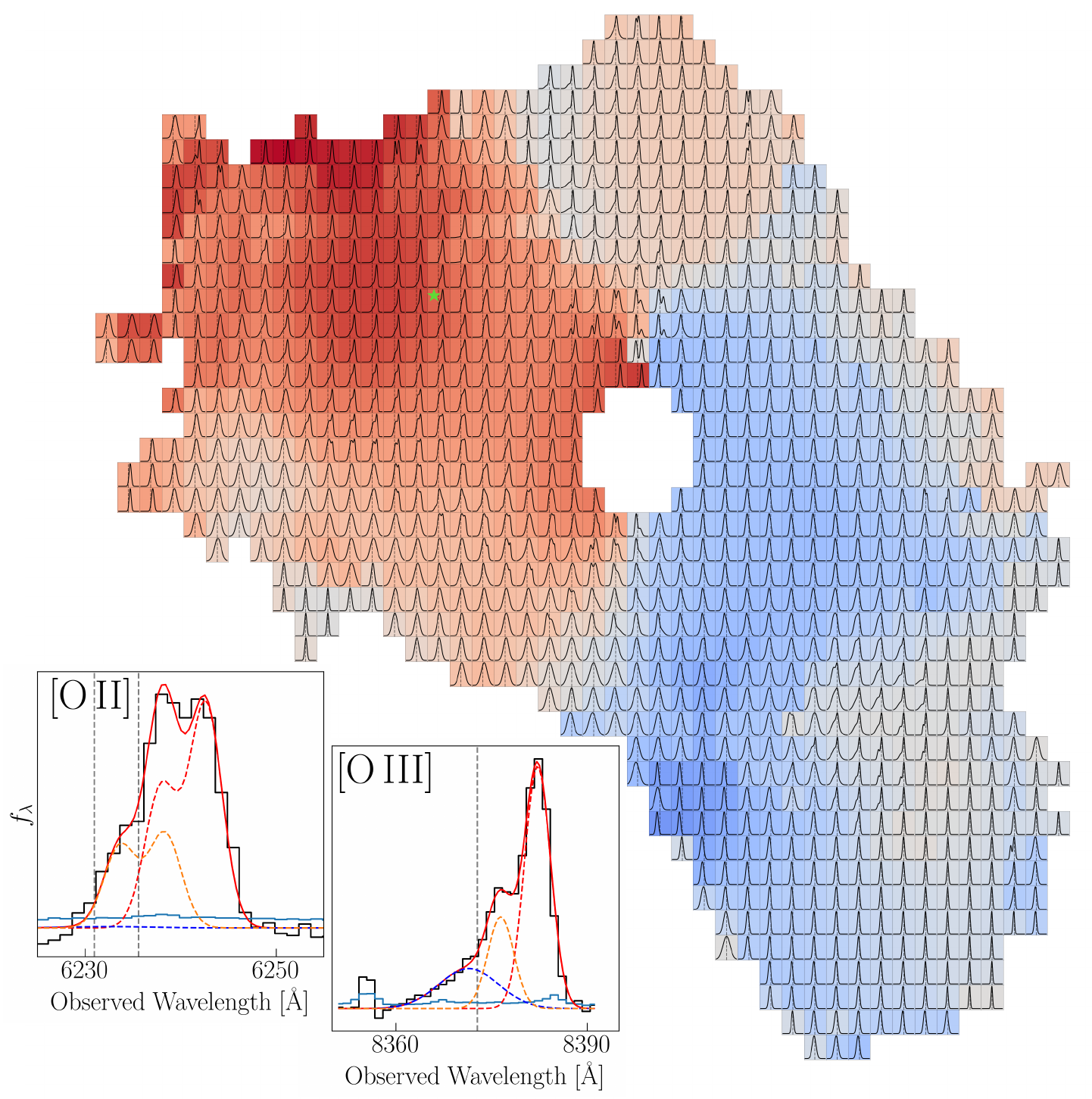}
    \caption{$\rm [O\,II]$ line profiles of the 3C\,57 nebula, with each cell corresponding to a spaxel in the MUSE datacube. Cells are colored based on their $\rm V_{50}$ values, and the fitted $\rm [O\,II]$ line profiles are shown in black, with a vertical line marking the quasar's systemic velocity. Profiles are normalized to unity, and each panel spans a velocity range of $\pm 800 \, \rm km\,s^{-1}$. This map was generated by first stacking the $\lambda3727$ and $\lambda3729$ components of each profile, and then show the summation of all profiles as a function of velocity. Two examples of line fitting for $\rm [O\,II]$ and $\rm [O\,III]$ in a spaxel are shown in the bottom-left, with its position on the nebula marked as a green star. The extracted spectrum and error array are shown as solid black and blue lines respectively. The best-fit models are shown as dashed red, orange, and blue lines respectively for each profile. For reference, $20$\,\AA \, in $\rm [O\,II]$ and $30$\,\AA \, in $\rm [O\,III]$ respectively corresponds to $1000 \, \rm km\,s^{-1}$. Multi-component emission features are seen across both the blueshifted and redshifted sides of the nebula, with a more prominent presence on the redshifted side. \textcolor{black}{In most spaxels with multi-component profiles, the individual components are blended, with centroid separations smaller than their respective widths, except in a few spaxels above the quasar centroid, likely due to residuals from quasar light subtraction.}}
    \label{fig:spec}
\end{figure*}

To better explore the complex kinematics of the nebula, we show a $\rm [O\,II]$ line profile map in Figure \ref{fig:spec}. We first stacked the $\lambda3727$ and $\lambda3729$ components of each kinematic component, and then summed all components in each spaxel to create the final map. Each cell corresponds to a spaxel in the MUSE datacube, and it is colored by the respective $\rm V_{50}$ value. This map allows us to visualize how line profiles correspond to the $\rm V_{50}$ in each spaxel and observe changes in the detailed line profile at different regions of the nebula. In particular, the nebula exhibits multi-component emission features on both the blueshifted and redshifted sides, with a stronger presence on the redshifted side. \textcolor{black}{In most spaxels with multi-component profiles, the individual components are blended, with centroid separations smaller than their respective widths, except in a few spaxels above the quasar centroid, likely due to residuals from quasar light subtraction.} Additionally, the northern knot of the nebula around G2 exhibits complex kinematics with a broad wing and a narrow core that we interpreted as the emission from the galaxy G2. In conclusion, the nebula displays complex emission features with broad emission wings.

\section{Discussion}
The discovery of a giant nebula requires both the presence of gas and its positioning within quasar's ionization cone. Gas arising from filamentary inflows, stellar and AGN feedback, galaxy interactions, and ISM/CGM gas of the host galaxy can create galactic-scale structures with regions ionized by the quasar. This gas will emit in recombination and collisionally-excited lines, producing giant quasar nebulae. Depending on the gas origin, quasar nebulae might exhibit distinct morphology, kinematics, and galactic environment. Specifically, the majority of the nebula around 3C\,57 is not morphologically or kinematically related to galaxies in the quasar host group, suggesting that it does not originate from interactions. Moreover, the nebula does not exhibit long filaments, which are expected from inflows. Instead, the nebula exhibits a blueshifted-redshifted pattern along the major axis, which can be produced by biconical outflows with proper alignment. The nebula also shows multi-component emission features with large dispersion. Such features are commonly observed around luminous quasars due to AGN feedback \citep[e.g.,][]{2012ApJ...746...86G, 2013MNRAS.430.2327L}. These features suggest that the 3C\,57 nebula may originate from AGN feedback. However, extended rotating gas can also produce a similar blueshifted-redshifted pattern. \textcolor{black}{Such extended rotating gas around massive ellipticals can originate from extended ISM/CGM disks that are kinematically aligned with the stellar component, or from late-stage mergers resulting in misalignment of the stellar and gas kinematics \citep{2012MNRAS.422.1835S}. In some cases, 21 cm emission around elliptical galaxies exhibits rotation-like kinematics but with circular velocity less than expected for such massive systems, suggesting the presence of net inflows or motion along elliptical orbits \citep{2001ApJ...562L..47F}. In these cases, the gas may have become part of the host galaxy and settled into regular rotation around the stellar body over several Gyr.} We will explore these possibilities in Sec \ref{sec:RD} and \ref{sec:AF}.

\subsection{Extended Rotating Gas}
\label{sec:RD}
One possibility is that the nebula primarily consists of rotating ISM and CGM gas from the quasar's host galaxy. If such gas is ionized by the quasar, it could produce a blueshifted-redshifted pattern along the major axis. To explore this, we compared the nebula's morphology and kinematics with H\,I gas detected around local early-type galaxies through 21 cm emission. H\,I 21 cm serves as an appropriate analog for the gas detected in optical emission around 3C\,57, which may have been neutral in the absence of ionizing photons from the luminous quasar.

\begin{figure*}
    \centering
    \includegraphics[scale=0.7]{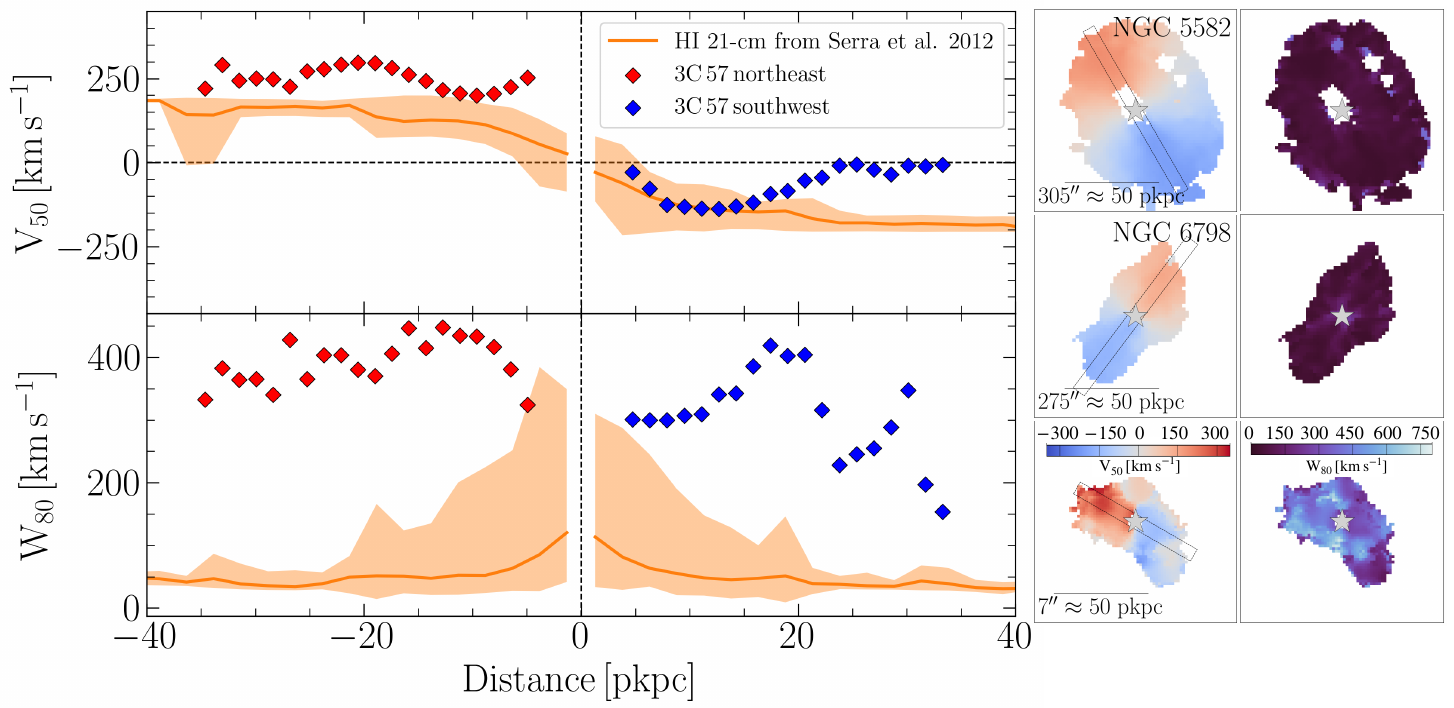}
    \caption{Left: Position-velocity and -velocity dispersion diagrams of 3C\,57 and H\,I gas in early-type galaxies \citep{2012MNRAS.422.1835S}. The radial profiles of $\rm V_{50}$ and $\rm W_{80}$ are shown respectively in the top and bottom panel. The binned 3C\,57 profiles are shown as red and blue diamonds representing northeast and southwest portion of the nebula. The mean radial profile of early-type galaxies is shown as a solid line, and the filled orange region marks the minimum and maximum value in each bin. Right: $\rm V_{50}$ and $\rm W_{80}$ maps of NGC\,6798, NGC\,5582, and 3C\,57 with the same colormaps shown in Figure \ref{fig:SBs}. The dashed rectangles in the left column represent the extraction slits with a width of $7 \rm \, kpc$, used to obtain the radial profiles shown on the left. 3C\,57 resembles early-type galaxies in its $\rm V_{50}$ map and profile, while the $\rm W_{80}$ shows significant discrepancies, suggesting the nebula around 3C\,57 is not typical extended rotating gas from a galaxy.}
    \label{fig:vprofile}
\end{figure*}

We used data from the $\rm ATLAS^{3D}$ H\,I survey, which contains 21-cm observations of 166 early-type galaxies in the local universe \citep{2011MNRAS.413..813C, 2012MNRAS.422.1835S}. From this sample, we selected galaxies classified as having large H\,I discs/rings by \citet{2012MNRAS.422.1835S}, characterized by regular rotation and a distribution extending beyond the galaxy's stellar body. We further refined our selection to include galaxies with gas disks comparable in size ($40{-}100 \, \rm kpc$) to the nebula around 3C\,57, excluding those with clear spiral structures. Our final set of galaxies includes NGC\,2685, NGC\,3941, NGC\,3945, NGC\,4262, NGC\,5582, NGC\,6798, and UGC\,06176 \citep{2006MNRAS.371..157M, 2009A&A...494..489J, 2010MNRAS.409..500O, 2012MNRAS.422.1835S, 2014MNRAS.444.3388S}. \textcolor{black}{The stellar masses of these galaxies range from $\log(M_{*}/\rm M_{\odot}) \approx 10.3$ to $\approx 11.0$, derived from detailed axisymmetric dynamical modeling \citep{2013MNRAS.432.1709C}.} To produce the $\rm V_{50}$ and $\rm W_{80}$ maps, we measured the velocity dispersion of H\,I in these galaxies by fitting the datacube with a single Gaussian, using the $\rm ATLAS^{3D}$ H\,I velocity maps as initial guesses. We then extracted kinematic profiles of the $\rm V_{50}$ and $\rm W_{80}$ maps using a pseudo-slit placed along the major axis, with a fixed width of $7\,\rm kpc$ and length of $100\,\rm kpc$. To facilitate comparison, we stacked the radial profiles of all galaxies and computed a mean profile, bounded by the minimum and maximum values. Figure \ref{fig:vprofile} displays this stacked profile alongside example $\rm V_{50}$ and $\rm W_{80}$ maps, in comparison to 3C\,57.

H\,I gas in early-type galaxies exhibits some similarities in amplitude and pattern in $\rm V_{50}$ maps and profiles to the 3C\,57 nebula (See Figure \ref{fig:vprofile}), although there are notable differences. An asymmetry is observed between the blueshifted and redshifted sides, with the blueshifted side approaches zero velocity starting at $15\, \rm kpc$, while the redshifted side reaching an amplitude of $250 \, \rm km\,s^{-1}$. This asymmetry arises from the presence of a luminous redward component on the blueshifted side and an extremely redward component on the redshifted side, which shift the velocities towards their corresponding values. The maximum velocity of the redshifted side is $\approx +250  \, \rm km\,s^{-1}$ while that of the blueshifted side is $\approx -150  \, \rm km\,s^{-1}$, representing a non-negligible asymmetry that exceeds the expected uncertainty in the quasar systemic redshift. Additionally, they differ significantly in velocity dispersion. The H\,I gas typically has a $\rm W_{80}\approx 50\,\rm km\,s^{-1}$, in contrast to the nebula's $\rm W_{80}\approx 300{-}400\,\rm km\,s^{-1}$. \textcolor{black}{We note that the kinematically quiescent region near the southwestern edge of the nebula exhibits $\rm W_{80}\approx 150\,\rm km\,s^{-1}$, which is lower than the rest of the nebula and causes the $\rm W_{80}$ to plateau at $+30 \, \rm kpc$, but it is still elevated compared to the H\,I gas detected in 21-cm around elliptical galaxies.} Moreover, the extended H\,I gas generally lacks multi-component emission, unlike 3C\,57. The kinematic inconsistencies between the nebula and the H\,I gas suggest that the 3C\,57 nebula might not consist solely of rotating ISM/CGM gas from the host galaxy. Instead, the gas may be related to rotation, but also disturbed by another process such as AGN feedback.

\subsection{AGN Feedback}
\label{sec:AF}
AGN feedback presents another plausible explanation for the nebula's origin. In this scenario, outflowing gas powered by the AGN extends from the nucleus, often forming a biconical structure. Biconical outflows are commonly observed in local Seyfert galaxies and has been extensively modeled to study the kinematics of the NLR in both Seyferts \citep[e.g.,][]{2000ApJ...532..247C, 2001AJ....121..198V} and quasars \citep[e.g.,][]{2014ApJ...786....3S}. Moreover, these models effectively reproduce various observed phenomena, including position-velocity diagrams \citep{2000ApJ...532L.101C} and line profiles \citep{2016ApJ...828...97B, 2023A&A...677A..58M}. When inclined to the line of sight, biconical outflows can produce a symmetric blueshifted-redshifted pattern, multi-component emission features, and large velocity dispersion. These characteristics align well with our observations of the 3C\,57 nebula, making the AGN feedback scenario a strong contender for explaining its origin and properties.

\begin{figure*}
    \centering
    \includegraphics[scale=0.7]{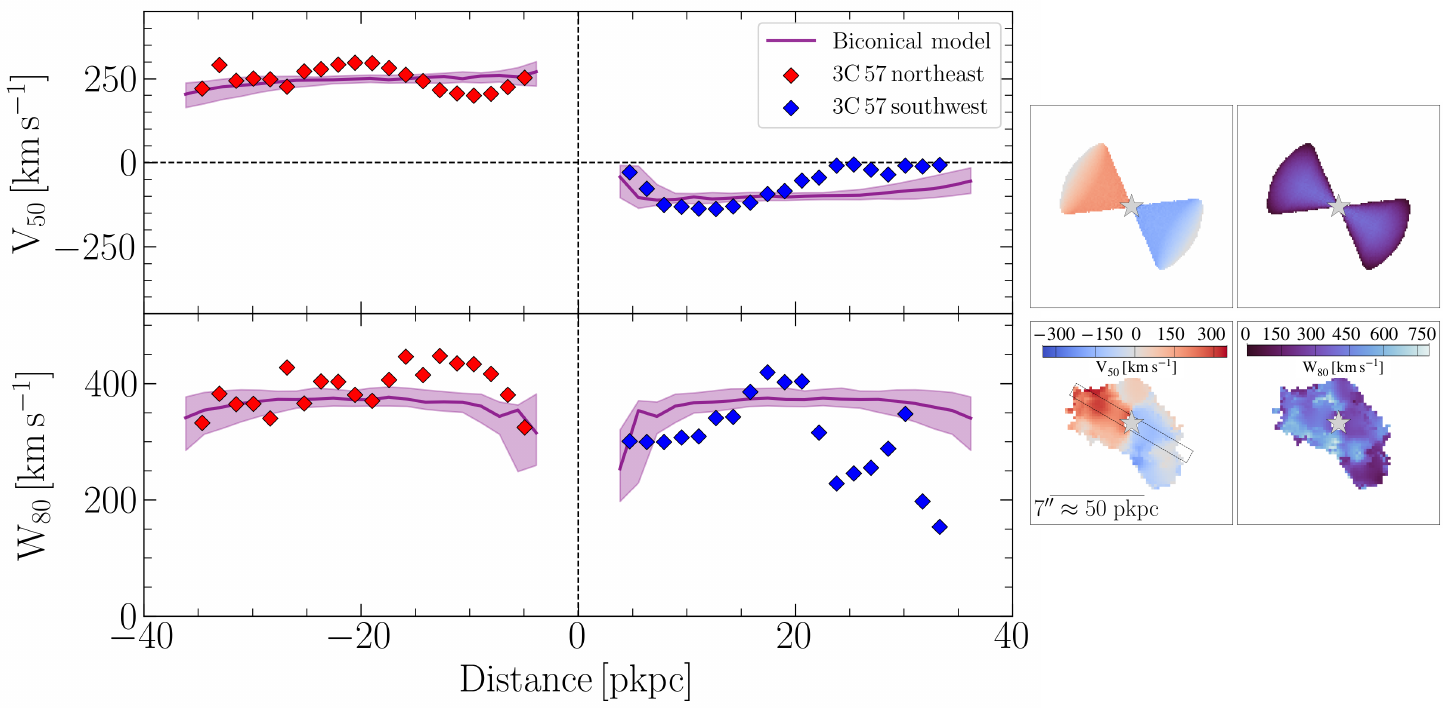}
    \caption{Same as Figure \ref{fig:vprofile} but for biconical models. The median posterior radial profile of the biconical models is shown as solid lines, and the filled purple region marks the minimum and maximum value in each bin. Right: $\rm V_{50}$ and $\rm W_{80}$ maps of the best-fit biconical model. Biconical outflow models can qualitatively reproduce the observed position-velocity and -velocity dispersion diagrams. However, the inferred $10{-}20^{\circ}$ inclination angle conflicts with the unobscured nature of the quasar, as the dusty torus is expected to be perpendicular to the outflow. Additionally, the nebula exhibits notably stronger emission along the minor axis compared to the model predication, underscoring the limitations of simple geometric models in fully capturing the complex nature of quasar nebulae. }
    \label{fig:bicprofile}
\end{figure*}

To determine whether a biconical model can reproduce the kinematics, we generated models following \citet{2016ApJ...828...97B} and \citet{ 2023A&A...677A..58M} and performed Markov chain Monte Carlo (MCMC) analysis. Each model is parameterized by three rotation angles, a velocity profile, semi-inner and -outer opening angles ($\theta_{\rm in}$, $\theta_{\rm out}$), and a flux profile. Due to the bicone's axial symmetry and the ability to rotate the model based on 3C\,57's position angle, we reduced three rotation angles to a single inclination angle. \textcolor{black}{The inclination is defined such that an edge-on bicone corresponds to zero inclination, while a face-on scenario results in a single visible cone projected as a circular region.} To create the relatively flat $\rm V_{50}$ and $\rm W_{80}$ profile, we adopted a constant velocity profile and a slow-decaying flux profile $f(r)=f_{0}e^{-\frac{r}{r_{0}}}$, where $f_0$ is the apex flux and $r_0$ is a variable scale-radius. \textcolor{black}{Based on experimentation, we assume a non-hollow geometry ($\theta_{\rm in}=0$), as the observed line profile and position-velocity profile are inconsistent with hollow models.} Our final model is characterized by three parameters: inclination angle ($\beta$), constant outflow velocity ($\rm v_{\rm out}$), and semi-outer opening angle ($\theta_{\rm out}$). 

We generated models within parameter grids of $5^{\circ}< \beta < 50^{\circ}$, $350 \, \rm km\, s^{-1} < v_{\rm out} < 1500 \, \rm km\, s^{-1}$, and $10^{\circ} < \theta_{\rm out} < 50^{\circ}$, following the ranges used in \citet{2016ApJ...828...97B}, but extend to greater values to accommodate potential model variations. We then extracted and interpolated the velocity and velocity dispersion profiles along the bicone's major axis using \texttt{RegularGridInterpolator} function from \texttt{scipy}. Additionally, we introduced a free parameter to allow for the velocity shift between the outflow and the measured quasar redshift. Finally, we ran MCMC with \texttt{emcee} \citep{2013PASP..125..306F} to estimate posteriors given the observed $\rm V_{50}$ and $\rm W_{80}$ profiles. During the minimization process, we imposed a velocity error floor of $20 \, \rm km \, s^{-1}$ to account for systematic errors from kinematics fitting.

The biconical outflow models can qualitatively reproduce the observed kinematics, and the resulting posteriors indicate a range of viable configurations. The parameter ranges that yield likely models are $13^{\circ} <\beta< 22^{\circ}$, $350 \, \rm km \, s^{-1} < v_{\rm max} < 720 \, km \, s^{-1}$, and $21^{\circ} < \theta_{\rm out}< 43^{\circ}$. Among these, the best-fit model has parameters of $\beta=20^{\circ}$, $v_{\rm max}=460 \, \rm km\, s^{-1}$, and $\theta_{\rm out}=34^{\circ}$. Figure \ref{fig:bicprofile} displays the PV diagrams from these posteriors, with the best-fit model shown as a purple line and the region between minimum and maximum values from all posteriors shaded. While the biconical models capture the overall kinematics, they fail to quantitatively reproduce certain features, such as the gradual $\rm V_{50}$ drop and $\rm W_{80}$ fluctuation at 20 kpc on the blueshifted side.

\subsection{An Ambiguous Origin}
Both extended rotating gas and biconical outflow models can qualitatively reproduce the blueshifted-redshifted pattern and velocity profile observed in the 3C\,57 nebula. However, unlike the 3C\,57 nebula, the H\,I gas in the extended rotating gas scenerio exhibits quiescent kinematics and does not display elevated velocity dispersion. While biconical models can qualitatively reproduce the observed velocity and velocity dispersion values, the small inclination angle derived from fitting the biconical outflow model is in tension with the unobscured nature of the quasar. Biconical outflows are more often observed in obscured AGN \citep{1996ApJ...463..498S}, which is expected from the unified model for AGN, where the dusty torus is perpendicular to the outflowing cones \citep[e.g.,][]{1995PASP..107..803U}. Therefore, with an inclination of $10{-}20^{\circ}$, we would expect an obscured quasar, which is inconsistent with the observed characteristics of 3C\,57. Additionally, we expect the line profiles from a biconical outflow to exhibit symmetry between the blueshifted and redshifted sides. In particular, the models predict that emission profiles on the blueshifted side will exhibit a lower surface brightness redward wing, and a similar blueward wing on the redshift side. However, the data reveal luminous blueshifted wings across nearly the entire nebula (see Figure \ref{fig:spec}). Moreover, the strong emission along the minor axis is not reproduced by the biconical models. These discrepancies highlight the limitations of simple geometric models in fully capturing the complex nature of quasar nebulae. Together, these findings suggest that neither model alone can fully explain the observed phenomena, indicating a more intricate interplay between the nebula's kinematics and its underlying processes.

Alternatively, the 3C\,57 nebula may originate from a combination of rotating gas and AGN feedback. In this scenario, the blueshifted-redshifted pattern could result from the rotating gas of the quasar's host galaxy, while the multi-component emission features with elevated dispersion may be induced by feedback processes. Similarly, elevated dispersion in molecular gas can be found in post-starburst galaxies, where strong turbulence is seen along with regularly rotating systems \citep[e.g.,][]{2022ApJ...929..154S, 2023ApJ...942...25F}. This scenario implies a dynamic interplay between the host galaxy's ISM/CGM and the AGN's energetic output. Such a model would explain the coexistence of the blueshifted-redshifted pattern and elevated dispersion, as well as the difficulty in fitting the data to simpler, single-origin models.

\subsection{Comparisons With Other Nebulae}
While both rotating gas and biconical outflow models cannot fully reproduce all observed features, we expanded our analysis by comparing the 3C\,57 nebula with other nebulae around UV luminous quasars reported in the CUBS+MUSEQuBES survey. Within this survey, 3C\,57 stands out as one of the nebulae with the highest velocity dispersions, alongside TXS\,0206${-}$048 and J\,2135${-}$5316. TXS\,0206${-}$048 showcases $\rm W_{80} \approx 450 \rm \, km \, s^{-1}$, possibly related to strong turbulence \citep{2024ApJ...962...98C}, while J\,2135${-}$5316 has $\rm W_{80} \approx 425 \rm \, km \, s^{-1}$ and exhibits complex multi-component emission features, possibly due to AGN feedback. In comparison, nebulae arising from interactions generally have lower dispersions, with $\sigma \approx 60{-}130 \rm \, km \, s^{-1}$ ($\rm W_{80} \approx 150{-}325 \, km \, s^{-1}$) (See Figure 8 of \citealt{2024ApJ...962...98C}, also see \citealt{2018ApJ...869L...1J, 2022ApJ...940L..40J, 2021MNRAS.505.5497H, 2024MNRAS.527.5429L}). None of the reported nebulae with elevated dispersion exhibit the distinct blueshifted-redshifted pattern seen in 3C\,57. In future work, we will provide a comprehensive characterization of giant nebulae in the CUBS+MUSEQuBES survey, with the potential to identify more samples of nebulae exhibiting blueshifted-redshifted features and gain deeper insights into their origins.

Additionally, the brightest cluster galaxies (BCGs) in cool-core clusters are useful analogs for understanding the origin of the 3C\,57 nebula. In these clusters, cool gas around BCGs is heated by energetic AGN feedback, with observable radio jets and X-ray cavities. $\rm [O\,II]$ nebulae near the positions of radio lobes and X-ray cavities have a velocity dispersion of $\sigma > 200 \, \rm km\,s^{-1}$, or equivalently $\rm W_{80}>500\, km\,s^{-1}$ \citep{2024ApJ...977..159G}. Additionally, elevated dispersions are observed in regions away from cavities, often in a direction roughly perpendicular to them in cool-core clusters. These features may indicate strong bipolar outflows, potentially from jets that are either unseen or have changed orientation over time \citep{2024ApJ...977..159G}. In the case of 3C\,57, two radio lobes are present: one centered near the quasar and another located $15''$ southeast (See Figure \ref{fig:SBs}). The 3C\,57 nebula does not extend to the southeast radio lobe, but is overlapped with the other radio lobe near the quasar's centroid. One possibility is that this central radio lobe indicates the presence of a jet impacting the nearby CGM and ISM, potentially accounting for the elevated dispersion observed in the nebula. Alternatively, the lobe may result from strong nuclear radio emission rather than a jet, with feedback that are either undetected or have shifted orientation over time. Regardless, 3C\,57 could serve as an intriguing low-mass analog to cool-core clusters, providing a unique perspective on how AGN-driven feedback and outflows might operate in lower-mass environments.

\section{Summary and Conclusions}
In this paper, we presented a case study of a giant nebula around a radio-loud quasar at  $z\approx0.672$ based on MUSE observations of the field of 3C\,57. We found that the giant, $\approx 70 \, \rm kpc$-scale nebula around 3C\,57 exhibits a blueshifted-redshifted pattern with large velocity dispersion and multi-component emission features. The observed active kinematics and blueshifted-redshifted morphology suggest that the nebula is unlikely to be solely rotating ISM/CGM gas from the host galaxy. Instead, the nebula's characteristics indicate a more complex origin, potentially resulting from a combination of rotating gas and AGN feedback mechanisms. 

While the 3C\,57 nebula provides a unique case study of the intricate interplay between AGN activity and host galaxy kinematics, our work will expand to analyze other nebulae in the CUBS+MUSEQuBES survey to determine if any exhibit a similar blueshifted-redshifted pattern. Insights from the 3C\,57 nebula may also shed light to the studies of other giant nebulae around starburst galaxies and galaxy groups \citep[e.g.,][]{2018A&A...609A..40E, 2019A&A...631A.114B, 2019ApJ...878L..33C, 2019Natur.574..643R, 2021MNRAS.507.4294Z, 2021ApJ...909..151B, 2022A&A...663A..11L, 2023MNRAS.522..535D, 2024A&A...683A.205E}. Future IFSs, such as LLAMAS \citep{2020SPIE11447E..0AF}, IFUM \citep{2022SPIE12184E..5PM}, Blue MUSE \citep{2019vltt.confE..24R}, and MIRMOS \citep{2020SPIE11447E..1EK}, will further drive discoveries of giant nebulae, and uncovering more systems like 3C\,57 to offer deeper insights.

\section{Acknowledgments}
We thank ZQL's thesis committee members, Camille Avestruz, Eric Bell, Joel Bregman, and Feige Wang for their valuable comments and discussions. SDJ and ZQL acknowledge partial support from HST-GO- 15280.009-A, HST-GO-15298.007-A, HST-GO-15655.018-A, and HST-GO-15935.021-A. SC gratefully acknowledges support from the European Research Council (ERC) under the European Union’s Horizon 2020 Research and Innovation programme grant agreement No 864361. JIL is supported by the Eric and Wendy Schmidt AI in Science Postdoctoral Fellowship, a Schmidt Futures program. MCC is supported by the Brinson Foundation through the Brinson Prize Fellowship Program. This paper is based on observations from the European Organization for Astronomical Research in the Southern Hemisphere under ESO (PI: J. Schaye, PID: 094.A-0131(B) \& 096.A-0222(A)), and the NASA/ESA Hubble Space Telescope (PI: L. Straka, PID: 14660). The reduced HST image can be found in MAST: \dataset[10.17909/x948-mv36]{http://dx.doi.org/10.17909/x948-mv36}. Additionally, this paper made use of the NASA/IPAC Extragalactic Database, the NASA Astrophysics Data System, \texttt{Astropy} \citep[][]{2022ApJ...935..167A}, \texttt{Aplpy} \citep{aplpy2012}, and \texttt{Photutils} \citep{larry_bradley_2023_7946442}.

\section*{DATA AVAILABILITY}
The data used in this paper are available from the the ESO and HST data archives.

\bibliography{main}{}
\bibliographystyle{aasjournal}



\end{document}